\documentclass[aps,prx,twocolumn,superscriptaddress,showpacs,longbibliography]{revtex4-1}

\usepackage{graphicx,color}

\definecolor{red}{rgb}{1,0,0}
\definecolor{blue}{rgb}{0,0,1}
\definecolor{green}{rgb}{0,1,0}
\definecolor{purple}{rgb}{0.7,0,1}

\usepackage{amsmath}

\usepackage{pdfpages}

\usepackage{hyperref}

\makeatletter
\AtBeginDocument{\let\LS@rot\@undefined}
\makeatother

\begin{document}

\preprint{APS/123-QED}

\title{Spin-dimer ground state driven by consecutive charge and orbital ordering transitions in the anionic mixed-valence compound Rb$_4$O$_6$}

\author{T. Knafli\v{c}}
\affiliation{Jo\v{z}ef Stefan Institute, Jamova c. 39, 1000 Ljubljana, Slovenia}

\author{P. Jegli\v{c}}
\affiliation{Jo\v{z}ef Stefan Institute, Jamova c. 39, 1000 Ljubljana, Slovenia}

\author{M. Komelj}
\affiliation{Jo\v{z}ef Stefan Institute, Jamova c. 39, 1000 Ljubljana, Slovenia}

\author{A. Zorko}
\affiliation{Jo\v{z}ef Stefan Institute, Jamova c. 39, 1000 Ljubljana, Slovenia}
\affiliation{Faculty of Mathematics and Physics, University of Ljubljana, Jadranska c. 19, 1000 Ljubljana, Slovenia}

\author{P. K. Biswas}
\affiliation{ISIS Pulsed Neutron and Muon Source, STFC Rutherford Appleton Laboratory, Didcot OX11 0QX, UK}

\author{A. N. Ponomaryov}
\affiliation{Dresden High Magnetic Field Laboratory (HLD-EMFL), Helmholtz-Zentrum Dresden-Rossendorf, 01328 Dresden, Germany}

\author{S. A. Zvyagin}
\affiliation{Dresden High Magnetic Field Laboratory (HLD-EMFL), Helmholtz-Zentrum Dresden-Rossendorf, 01328 Dresden, Germany}

\author{M. Reehuis}
\affiliation{Helmholtz-Zentrum Berlin f{\"u}r Materialien und Energie, 14109 Berlin, Germany}

\author{A. Hoser}
\affiliation{Helmholtz-Zentrum Berlin f{\"u}r Materialien und Energie, 14109 Berlin, Germany}

\author{M. Gei\ss }
\affiliation{Institute of Physical Chemistry and Center for Materials Research, Justus-Liebig-
University Giessen, Heinrich-Buff-Ring 17, 35392 Giessen, Germany}

\author{J. Janek}
\affiliation{Institute of Physical Chemistry and Center for Materials Research, Justus-Liebig-
University Giessen, Heinrich-Buff-Ring 17, 35392 Giessen, Germany}

\author{P. Adler}
\email{adler@cpfs.mpg.de}
\affiliation{Max Planck Institute for Chemical Physics of Solids, N{\"o}thnitzer Stra\ss e 40, 01187
Dresden, Germany}

\author{C. Felser}
\affiliation{Max Planck Institute for Chemical Physics of Solids, N{\"o}thnitzer Stra\ss e 40, 01187
Dresden, Germany}

\author{M. Jansen}
\email{m.jansen@fkf.mpg.de}
\affiliation{Max Planck Institute for Chemical Physics of Solids, N{\"o}thnitzer Stra\ss e 40, 01187
Dresden, Germany}

\author{D. Ar\v con}
\email{denis.arcon@ijs.si}
\affiliation{Jo\v{z}ef Stefan Institute, Jamova c. 39, 1000 Ljubljana, Slovenia}
\affiliation{Faculty of Mathematics and Physics, University of Ljubljana, Jadranska c. 19, 1000 Ljubljana, Slovenia}

\date{\today}

\begin{abstract}
{Recently, a Verwey-type transition in the mixed-valence alkali sesquioxide Cs$_4$O$_6$ was deduced from the charge ordering of molecular peroxide O$_2^{2-}$ and superoxide O$_2^-$ anions accompanied by the structural transformation and a dramatic change in electronic conductivity [Adler et al, Sci. Adv \textbf{4}, eaap7581 (2018)]. Here, we report that in the sister compound  Rb$_4$O$_6$ a similar Verwey-type charge ordering transition is strongly linked to O$_2^-$  orbital and spin dynamics. On cooling, a powder neutron diffraction experiment reveals  a charge ordering and a cubic-to-tetragonal transition at $T_{\rm CO}=290$~K, which is followed by a further structural instability at $T_{\rm s}=92$~K that involves an additional reorientation of magnetic O$_2^-$ anions. Magnetic resonance techniques supported by density functional theory computations suggest the emergence of a peculiar type of $\pi^*$-orbital ordering of the magnetically active O$_2^-$ units, which promotes the formation of a quantum spin state composed of weakly coupled spin dimers. These results reveal that similarly as in 3$d$ transition metal compounds, also in in the $\pi^*$ open-shell alkali sesquioxides the interplay between Jahn-Teller-like electron-lattice coupling and Kugel-Khomskii-type superexchange  determines the nature of orbital ordering and the magnetic ground state.}
\end{abstract}

%\pacs{76.60.-k, 75.50.Cc, 73.43.Nq, 74.70.Xa}

\maketitle
\section{Introduction}

Research of mixed-valence compounds has come a long way over the course of roughly  a century  - from observing and explaining the interesting color in Prussian Blue  \cite{robin1962color}, to studying charge ordering, giant magnetoresistance and superconductivity in various mixed-valence transition-metal compounds \cite{rao1998colossal,uehara1999percolative,chang2012direct}. An interplay of lattice, charge, orbital and spin degrees of freedom is a common thread in these systems and one notable example of such phenomena is the Verwey transition in Fe$_3$O$_4$ \cite{verwey1939electronic}. In this archetypal mixed-valence compound comprising  Fe$^{2+}$ and Fe$^{3+}$ ions at low temperatures, the structural transition characterized by charge ordering is accompanied by a significant drop in the sample conductivity and drastic change in the magnetic order. The Verwey transition is historically important, as it represents one of the first attempts to explain the metal to insulator transition in mixed-valence compounds \cite{verwey1947physical}. However, the actual low-temperature charge ordering pattern, probably coupled to the  structural fluctuations that persist well above the Verwey transition temperature,  has recently turned out to be much more complex than  originally anticipated by Verwey \cite{senn2012charge,perversi2019co}.

\begin{figure}[t]
\includegraphics[width=\linewidth]{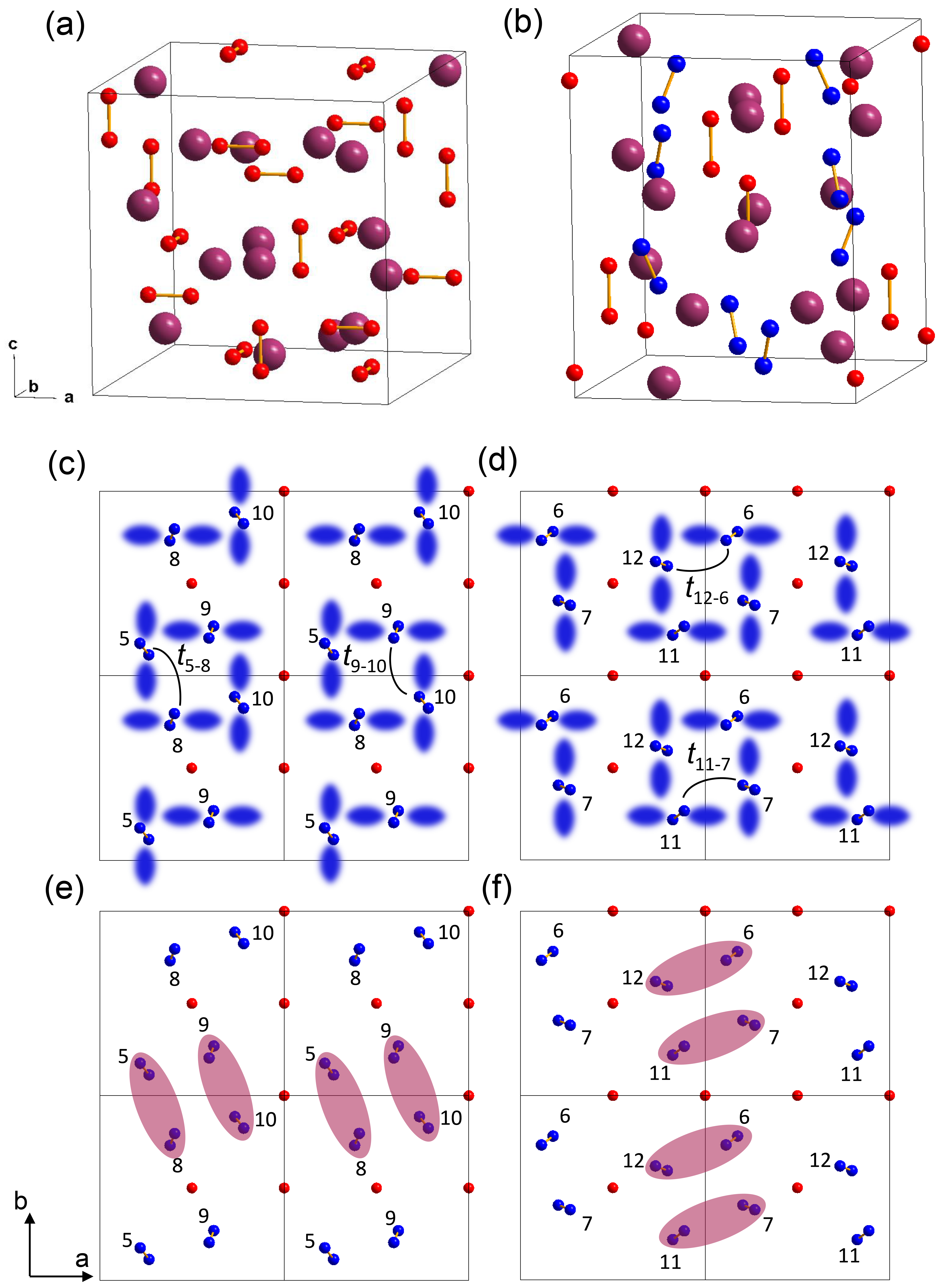}
\caption{Crystal structures of alkali sesquioxides $A_4$O$_6$ in (a)  the high-temperature charge-disordered cubic (space group $I\bar{4}3d$) and (b) in the charge-ordered tetragonal (space group $I\bar{4}$) phase. (c-d) The low-temperature tetragonal phase of Rb$_4$O$_6$ (space group $P\bar{4}$)  is characterized by an additional tilt of magnetic O$_2^-$ units promoting their alternating $\pi^*$-orbital ordering (molecular orbital lobes of different O$_2^-$ units labeled from 5 to 12 are indicated by  the blue shaded areas). The structure is shown in the tetragonal $a-b$ plane for slices $0\geq z\geq 0.5$ (c) and $0.5\geq z\geq 1$ (d). Solid black lines in (c) and (d) show the main inter-site hopping integrals $t_{i,j}$ between nearest neighboring O$_2^-$  $i$ and $j$ sites. The resulting formation of spin dimers depicted as pink ellipses  are presented for $0\geq z\geq 0.5$ (e) and for $0.5\geq z\geq 1$ (f). In all figures,  small blue spheres represent oxygen atoms in  O$_2^-$ units, small red spheres oxygen atoms in  O$_2^{2-}$ units and large violet spheres Rb atoms.}
\label{fig:structure}
\end{figure}

Alkali sesquioxides ($A_4$O$_6$, $A$ = Rb, Cs) are anionic mixed-valence compounds with simple cubic structure at room temperature [Fig. \ref{fig:structure}(a)]  and electronic states dominantly affected by $\pi$-molecular orbitals of O$_2$ units \citep{sans2014structural,jansen1999rb4o6,jansen1991neue,helms1939kristallstrukturen, arvcon2013influence}. The four alkali metals donate four electrons to the three O$_2$ molecules, resulting in their average charge state of O$_2^{-4/3}$. Physically, three electrons are localized each on it's own O$_2$ molecule, leaving the fourth one delocalized between structurally equivalent O$_2$ units.
A structural transition from the cubic to the tetragonal phase~[Fig. \ref{fig:structure}(b)] occurring in Cs$_4$O$_6$ at 250~K is accompanied by charge ordering \cite{adler2018verwey, colman2019} as the fourth electron orders in such a way that a mixed-valence state of magnetic O$_2^-$ and diamagnetic O$_2^{2-}$ anions is formed in the ratio of 2:1. As a result, the conductivity drops by several orders of magnitude \cite{adler2018verwey}, bearing striking similarities with the Verwey transition \cite{verwey1939electronic}.
The molecular nature of charge states allows charge dynamics to be followed  by various spectroscopic techniques, such as Raman and impedance spectroscopy,  showing that charge fluctuations happen in the cubic phase on a time scale of $10^{-12}$~s. Electron paramagnetic resonance (EPR) and nuclear magnetic resonance (NMR) techniques, on the other hand, demonstrate full charge localization in the tetragonal phase, on their time scales of $10^{-10}$~s and $10^{-6}$~s, respectively.

Obviously, the charge ordering in Cs$_4$O$_6$ has features in common with charge-ordering phenomena in transition metal compounds like Fe$_3$O$_4$ or the manganites. In fact, analogies in the physical properties of the open $p$ shell mixed-valence alkali sesquioxides as well as of the single-valence superoxides having merely magnetic O$_2^-$ ions in  their lattice with those of 3$d$ transition metal compounds make them an attractive class of compounds for studying strongly correlated electron physics. Here, the lattice degree of freedom is determined by the tilting of the O$_2^-$ units, which similarly as the Jahn-Teller effect in 3$d$ systems, removes the degeneracy of the highest occupied $\pi^*$ molecular orbitals and  may give rise to orbital ordering.
For instance, in CsO$_2$ orbital ordering is suggested to drive the formation of quasi-one-dimensional spin chains \cite{riyadi2012antiferromagnetic, klanjvsek2015phonon, Kanflic2015}, in analogy to  KCuF$_3$ \cite{kadota67,pavarini08}, whereas RbO$_2$ does not show any signatures of low-dimensional magnetism \cite{astuti19}. The differences in the magnetic properties of the superoxides were attributed to differences in the low-temperature crystal structures leading to different tilting and thus orbital ordering patterns for the molecular O$_2^-$ units \citep{astuti19,kim14}.

While the charge and lattice dynamics are now well understood in Cs$_4$O$_6$, it remains unclear how the spin and orbital degrees of freedom in the mixed-valence sesquioxides respond to the Verwey-like charge ordering and whether similar structural and orbital fluctuations as in Fe$_3$O$_4$ exist and affect the low-temperature electronic state as well. Moreover, the nearly degenerate $\pi^*$ strongly correlated states have to be treated explicitly in any realistic model where orbital degrees of freedom are included  on equal footing with the electron spins, probably even involving Kugel-Khomskii-type superexchange \cite{kim14,kugel73,oles2017orbital}. Earlier theoretical calculations for $A_4$O$_6$ neglected such aspects and suggested a highly degenerate magnetic ground state, comprising of many competing noncollinear spin configurations \citep{winterlik2009challenge}.
In the present work we demonstrate that  the magnetic properties of the sesquioxides $A_4$O$_6$ sensitively depend on the alkali cation $A$ and that the differences in the magnetic properties can be traced back to the important coupling to $\pi^*$ orbital degrees of freedom. We first show, that  Rb$_4$O$_6$ at 290~K  undergoes a Verwey-type charge-ordering transition from the cubic to a tetragonal structure of space group $I\bar{4}$, which is the same as in Cs$_4$O$_6$. In addition, however, Rb$_4$O$_6$ shows a further structural transition, which incorporates a peculiar type of $\pi^*$ long-range orbital ordering below $T_{\rm s}=92$~K and promotes a strong antiferromagnetic coupling between pairs of O$_2^-$ $S=1/2$ spins. Thus, a magnetic ground state of weakly coupled spin dimers emerges in Rb$_4$O$_6$ within the charge and orbitally ordered state.
Our results indicate that the orbital ordering in Rb$_4$O$_6$ at 92~K may be driven by exchange interactions (Kugel-Khomskii-type superexchange \cite{kugel73}) and demonstrate the strongly intertwined charge, orbital, spin and lattice degrees  of freedom. These findings further extend the analogies between the physical phenomena of open-shell-$p$-electron systems and 3$d$ compounds  to mixed-valence alkali sesquioxides.

\section{Results}

\subsection{Charge ordering - Verwey transition}

\begin{figure}[t]
\includegraphics[width=\linewidth]{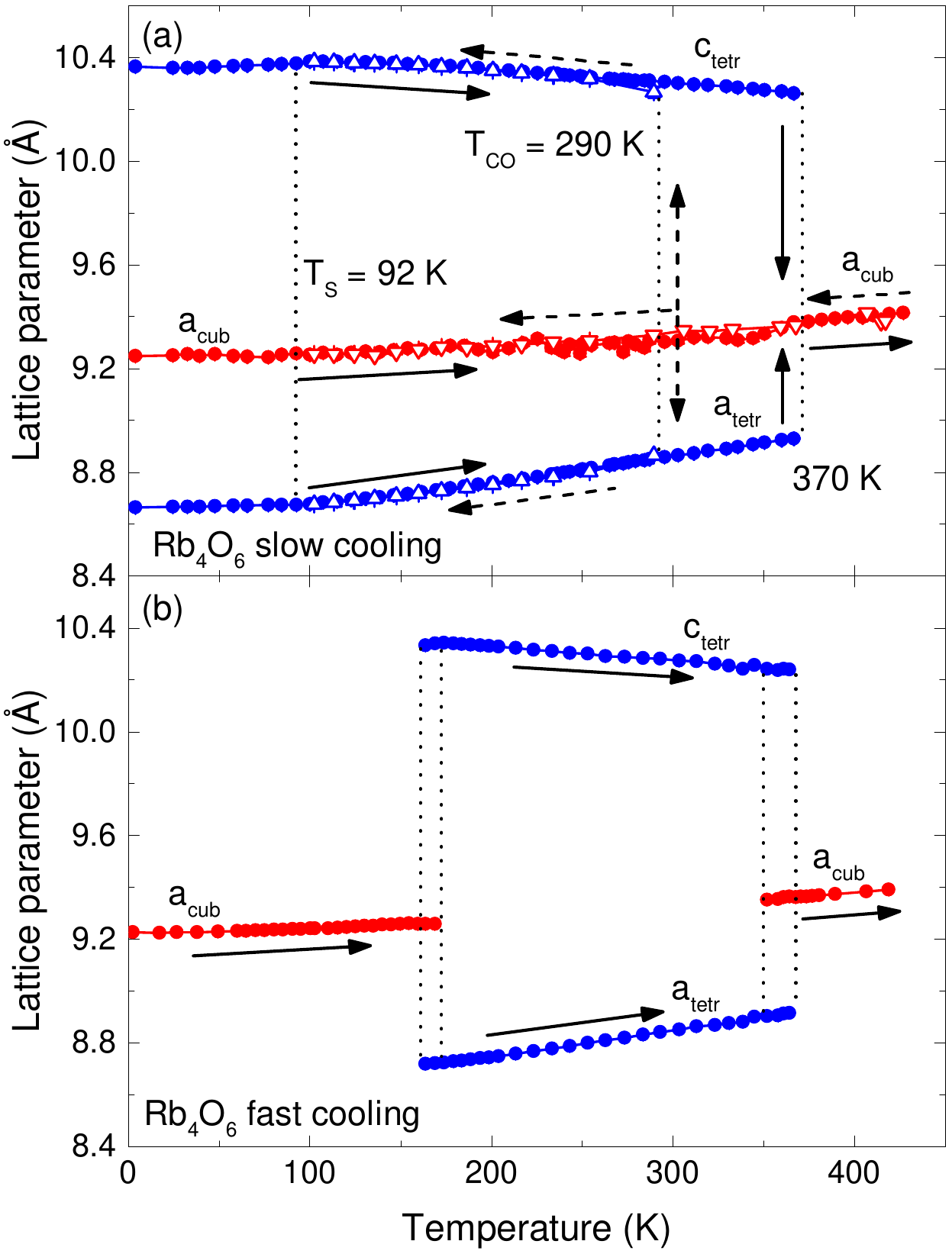}
\caption{The temperature dependencies of the lattice parameters for the cubic (red) and the tetragonal (blue) phases of Rb$_4$O$_6$. (a) The temperature dependence of these parameters during the slow cooling (open triangles) and subsequent heating (circles) experiments. (b) Temperature dependencies of the lattice parameters for the measurements taken on warming  after the quench-cooling thermal protocol. Dashed and solid arrows indicate cooling and heating temperature protocols respectively and the transformations from the cubic to the tetragonal and back to the cubic phases.}
\label{fig:lattice}
\end{figure}

Powder neutron diffraction (PND) studies verify that Rb$_4$O$_6$, similar as its sister compound Cs$_4$O$_6$ \cite{adler2018verwey}, undergoes a structural transition from the cubic (space group $I\bar{4}3d$, No. 220) to the tetragonal (space group $I\bar{4}$, No. 82) crystal structure, which mainly proceeds between 290 and 250 K (Figs. S1 and S2 in Ref. \cite{supplemental}). Even when  an extremely slow cooling procedure is applied a fraction of 2.2\% of cubic phase still persists at 2~K. The incomplete transformation reflects the kinetic hindrance of the transition, which involves a large reorientation of the molecular O$_2$ units. While in the cubic structure [Fig. \ref{fig:structure}(a)] the O$_2$ anions are oriented perpendicular to each other along  the three equivalent crystal axes, in the tetragonal structure the diamagnetic O$_2^{2-}$ align along thetetragonal $c$-axis whereas the two paramagnetic O$_2^-$ ions per formula unit tilt by an angle $\beta$ with respect to the  $c$-axis. The large molecular reorientation, which is necessary to stabilize the tetragonal structure,   gives rise to a first order structural transition with  a thermal hysteresis width of $\sim$~100~K [Fig.\ref{fig:lattice}~(a)] and most probably is also the main reason for the previously reported history dependence of the magneto-structural properties of the sesquioxides \cite{arvcon2013influence}. In fact, in a rapid quench-cooling experiment we are able to completely freeze-in the ambient-temperature cubic phase (Fig. S3 in Ref. \cite{supplemental}), which on warming remains metastable up to $\sim 160$~K, where it transforms into the more stable tetragonal phase (Fig. \ref{fig:lattice}b). The tetragonal phase finally changes back into the high-temperature cubic phase in the temperature range between 350 and 370~K, irrespective of sample history.

Refinement of the high-resolution PND pattern collected at 400~K confirms the charge-disordered state of the cubic phase with peroxide and superoxide units occupying the same Wyckoff position in random distribution. The resulting O-O bond length amounts to 1.35 \AA . Refinement of the 100~K PND pattern reveals distinct O$_2^-$ and O$_2^{2-}$ anions with O-O bond lengths of 1.31 and 1.52~\AA ,  respectively, for the charge-ordered tetragonal structure (see Ref. \cite{supplemental} for refinement details). The O$_2^-$ tilt angle $\beta \sim 20^\circ$ is slightly larger than $\beta \sim 17^\circ$ for Cs$_4$O$_6$. The main difference to Cs$_4$O$_6$ is a smaller unit cell volume (788.4 \AA$^3$ and 920 \AA$^3$ for Rb$_4$O$_6$ and Cs$_4$O$_6$, respectively) as the ionic radius of Rb$^+$ is smaller than that of Cs$^+$, which has important implications for the structural and magnetic properties below 100~K as it will be described later.

\begin{figure}[t]
\includegraphics[width=\linewidth]{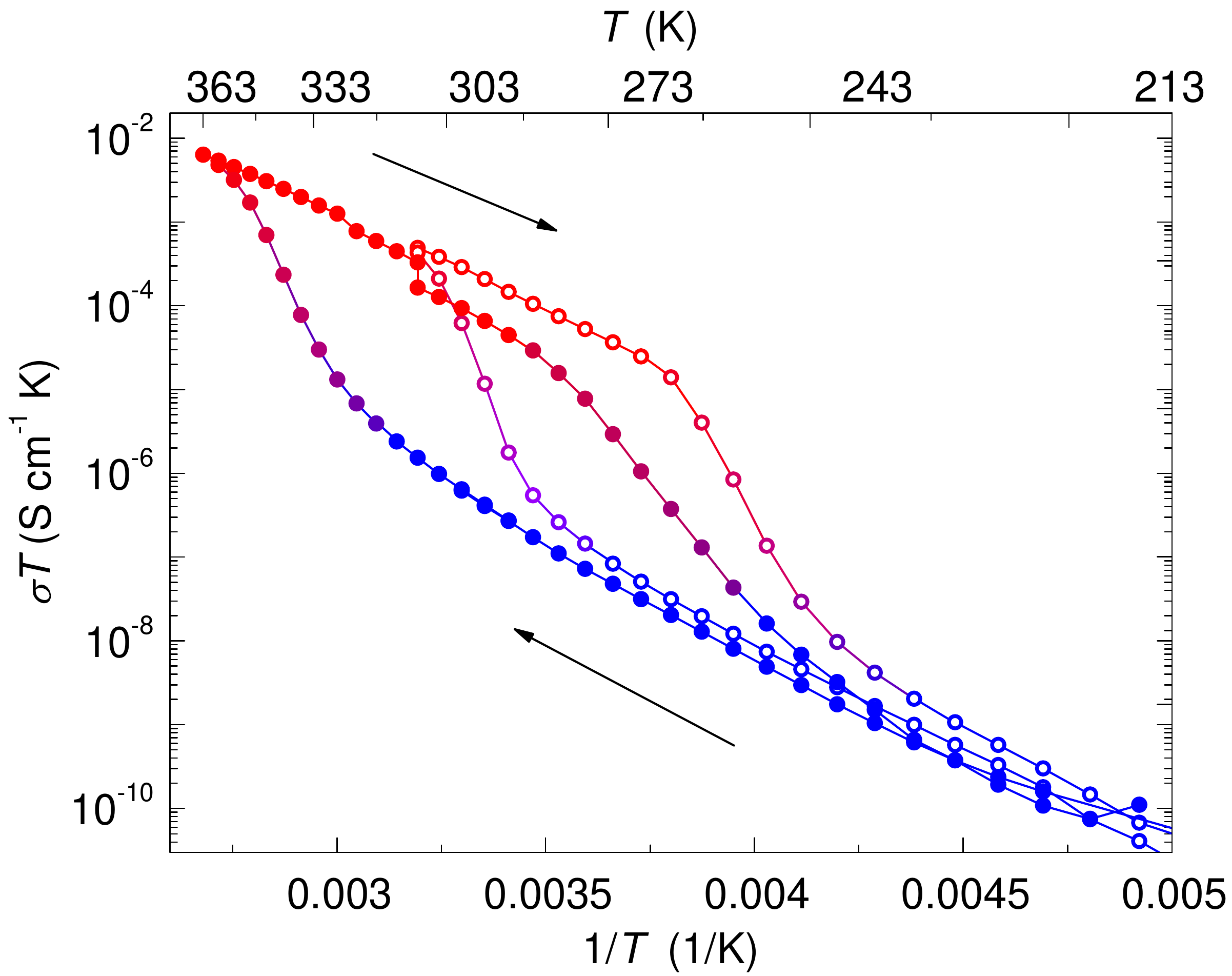}
\caption{Arrhenius plot of the conductivity $\sigma$ multiplied by $T$ for Rb$_4$O$_6$ (solid circles). For comparison,  the data for Cs$_4$O$_6$ (open circles) \cite{adler2018verwey} are shown. The color of symbols indicates the structure -- red symbols stand for the cubic phase and blue symbols for the tetragonal phase.}
\label{fig:conductivity}
\end{figure}

\begin{figure}[t]
\includegraphics[width=\linewidth]{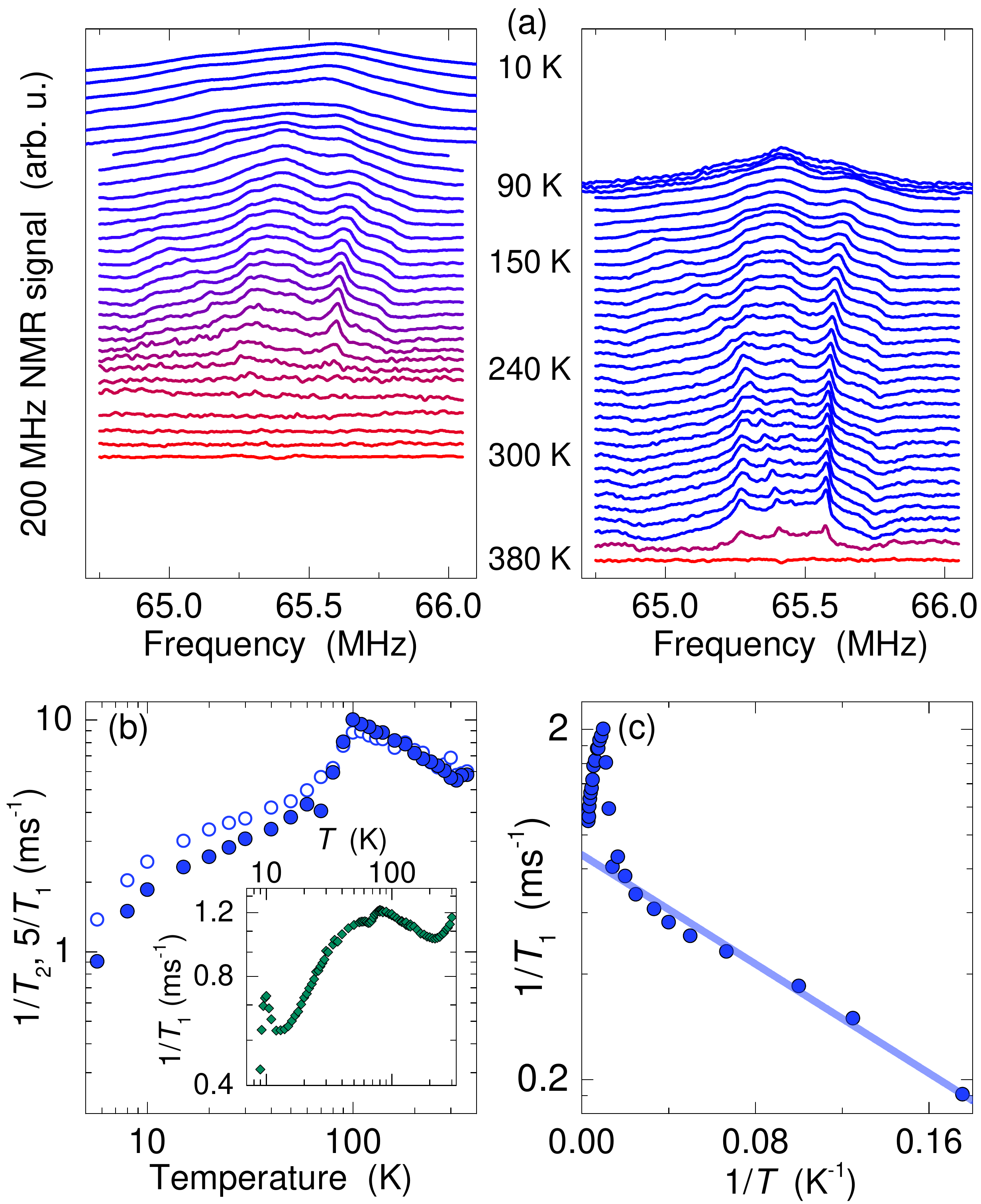}
\caption{ (a) The temperature dependence of $^{87}$Rb NMR spectra measured on cooling (left) and on heating (right). Red color of the spectra indicates the cubic phase while blue stands for the tetragonal phase. The horizontal base line for each spectrum inticate the temperature where it was measured. (b) The temperature dependencies of $5/T_1$ (solid circles) and $1/T_2$ (open circles) relaxation rates. In the inset, the $^{133}$Cs $1/T_1$ data of CsO$_2$ is shown for comparison (data taken from \cite{klanjvsek2015phonon}). (c) Solid circles show $1/T_1$ in the reciprocal $1/T$ scale. Linear regime of thermally activated behavior yielding a spin gap $\Delta _{\rm s}/k_{\rm B} =9$~K is fitted for $T\leq 70$~K (solid line).}
\label{fig:druga}
\end{figure}

\begin{figure*}[t]
\includegraphics[width=\linewidth]{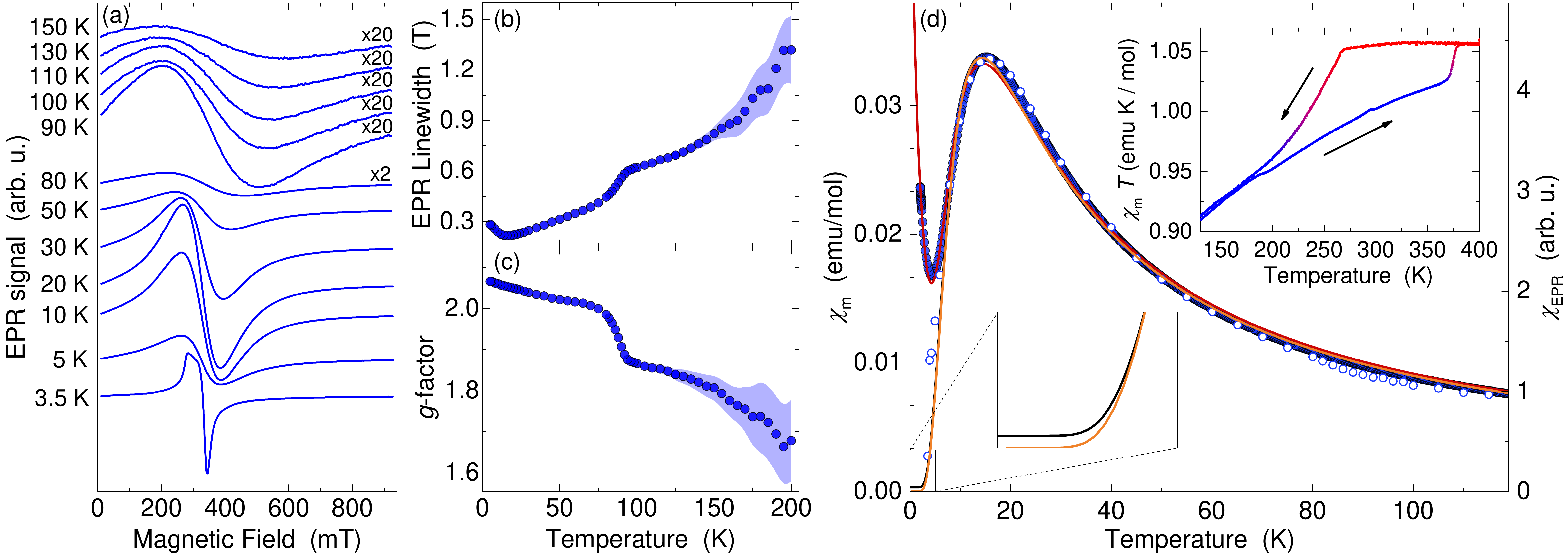}
\caption{(a) The temperature dependence of the X-band EPR signal. (b) The temperature dependence of EPR linewidth and (c) of \textit{g}-factor. The blue area at high temperatures indicates the level of uncertainty. (d) The temperature dependence of the molar spin susceptibility as obtained from bulk magnetization (solid circles) and EPR (open circles) measurements. The red line shows an isolated-dimer-model (Eq. \ref{BBeq}) fit to the magnetic susceptibility data with an added small paramagnetic contribution. The orange line is an isolated-dimer-model fit to EPR data with no paramagnetic contribution. The black line shows a small correction to the spin susceptibility  for the added DM interaction of 2.5~K (expanded low-temperature range is shown in the small inset). The inset presents the $\chi _{\rm m} T$ data revealing anomalies associated with the cubic-tetragonal and the reverse tetragonal-cubic transition.}
\label{fig:tretja}
\end{figure*}

The Verwey-type nature of the charge-ordering transition in Rb$_4$O$_6$, just as for Cs$_4$O$_6$~\cite{adler2018verwey}, is apparent from a change in the electronic conductivity (see Ref. \cite{supplemental} for details of impedance spectroscopy measurements) by two orders of magnitude in response of the structural transition (Fig. \ref{fig:conductivity}).  Local probe NMR and EPR techniques  also detect freezing of the charge dynamics at the cubic to tetragonal transition on their respective time-scales. $^{87}$Rb ($I=3/2$) NMR measurements on a pristine sample in the cubic phase show no signal at room temperature [Fig. \ref{fig:druga}(a)]. This is reminiscent of Cs$_4$O$_6$, where the absence of $^{133}$Cs and $^{17}$O NMR signals was attributed to strong relaxation effects in the cubic charge-disordered state \cite{adler2018verwey}. Namely, as the charge rapidly fluctuates on the NMR time-scale of $10^{-6}$~s between equivalent O$_2$ units, both quadrupole and hyperfine coupling interactions are strongly modulated, thus reducing the nuclear relaxation times. On slow cooling, the $^{87}$Rb NMR spectra start to emerge below 250~K. This marks the suppression of $^{87}$Rb relaxation rates due to the freezing of charge dynamics and the onset of charge ordering. Finally, in the X-band EPR measurements, the EPR signal appears on slow cooling as a very broad Lorentzian line below 250~K [Fig. \ref{fig:tretja}(a) and Fig. S4]. The appearance of the EPR signal provides additional evidence for the cubic-tetragonal structural and charge-ordering transition on the EPR time-scale of $10^{-10}$s. We note that in all these spectroscopic experiments, which are sensitive to the charge dynamics (i.e., impedance spectroscopy, EPR, and NMR), we find clear thermal hysteresis behavior [Figs. \ref{fig:conductivity} and \ref{fig:druga}(a)] of the Verwey-like charge  ordering transition.

Finally, in analogy to Cs$_4$O$_6$ \cite{adler2018verwey}, the structural/charge ordering phase transition is in Rb$_4$O$_6$ also  reflected as a small anomaly in the molar magnetic susceptibility $\chi_{\rm m}$($T$). This is particularly evident from the product $\chi_{\rm m} T$, which suddenly decreases below 270~K on slow cooling from the initial $T=400$~K and fully recovers back on warming at 375~K [inset to Fig. \ref{fig:tretja}(d)].

\subsection{The second structural transition - orbital ordering}

\begin{figure*}[htbp]
\includegraphics[width=0.9\textwidth]{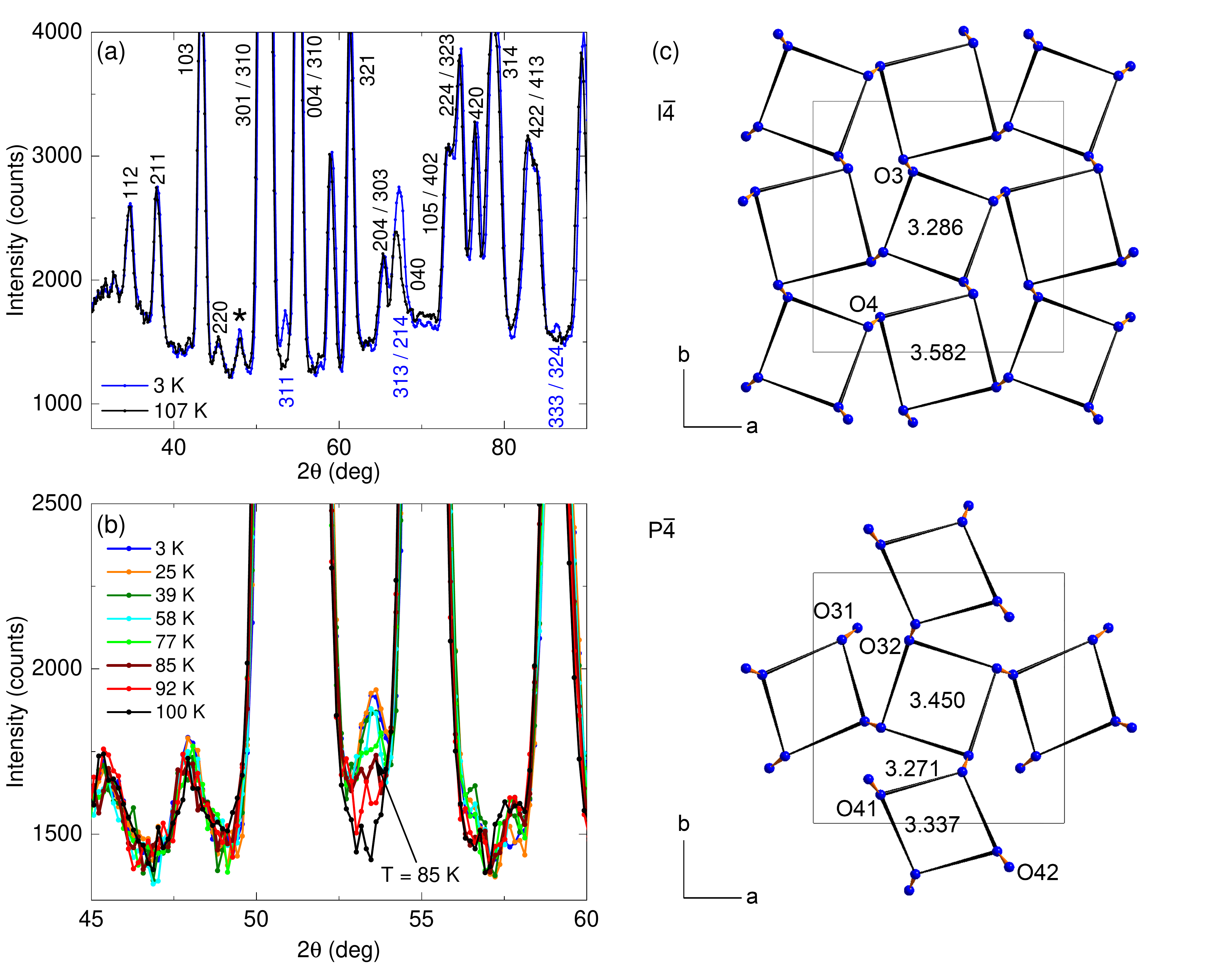}
\caption{(a) Powder neutron patterns of Rb$_4$O$_6$ collected on the instrument E6 ($\lambda = 2.426$~\AA ) at 3 and 107 K. At 3 K additional weak reflections have been detected, which can be indexed as primitive reflections following the extinction rule $h + k + l$ = odd (labeled in blue) and which are forbidden in the $I$-centered tetragonal structure obeying the rule $h + k + l$ = even (labeled in black). The asterisk indicates the main reflection of residual cubic phase. (b) Powder neutron  patterns in a reduced  $2\theta$ range showing the temperature dependence of the superstructure reflection 311. (c) Shortest intermolecular oxygen separations in the high-temperature $I\bar{4}$ (top) and the low-temperature $P\bar{4}$ (bottom) structure of Rb$_4$O$_6$. Atomic labels are as in Table S2 (the equivalent molecule labels as used in Fig. \ref{fig:structure} are given in Table S3), distances are in \AA. In $I\bar{4}$ all quadrilateral polygons are formed by symmetry equivalent oxygen atoms resulting in degenerate lengths of the edges and in a point group symmetry of $\bar{4}$ for both sets of polygons. In $P\bar{4}$ the coupling of the original O4 atoms due to space group symmetry is removed and the degeneracy of the intermolecular separations of O41 and O42 is lifted, while the sub-sets of O31 and O32 still form equilateral polygons of local $\bar{4}$ symmetry, although of substantially different lengths.}
\label{fig:super}
\end{figure*}

A further structural phase transition is for  Rb$_4$O$_6$ apparent from the emergence of additional Bragg reflections in the PND patterns  collected at $T\leq 92$~K [Fig. \ref{fig:super}(a) and (b)]. These reflections can be indexed as primitive reflections following the extinction rule $h + k + l$ = odd, if persistence of a tetragonal unit cell is assumed. To demonstrate this, we compare in Fig. \ref{fig:super}(a) the powder patterns collected at 3 and 107~K, where the presence of some prominent superstructure reflections is evident. The superstructure reflections appear  at $T_{\rm s}=92$~K, as exemplified for the 311 reflection in Fig. \ref{fig:super}(b).  It is remarkable that the intensities of the reflections of the $I$-centered structure ($h + k + l$ = even) remain practically unchanged. The violation of the extinction rule clearly indicates the loss of the $I$-centering and indeed, the high-resolution powder pattern of Rb$_4$O$_6$ collected at 3~K could be successfully refined in $P\bar{4}$ (space group No. 81), the maximal $klassengleiche$ subgroup of $I\bar{4}$ (space group No. 82), see Fig. S2 and Tables S1 and S2 in Ref. \cite{supplemental} for the refinement results and calculated structural parameters. 

The second structural  transition at $T_{\rm s}$ is also reflected in an anomaly in the lattice parameters [Fig. \ref{fig:lattice} (a)]. While above 100~K $a_{\rm tet}$ decreases and $c_{\rm tet}$ increases with decreasing temperature, an inflection point in $a_{\rm tet}(T)$ and a maximum in $c_{\rm tet}(T)$ is seen near 100~K. In contrast to the cubic-tetragonal transition, the volume change of the $I\bar{4}$ - $P\bar{4}$ transition ($\sim 5$~\AA$^3$) is small and the thermal contraction is virtually isotropic. More importantly, as a consequence of the loss of the $I$-centering in course of the transition, in the low-temperature crystal structure the O atoms of the two O$_2^-$ units are allowed to vary independently. One O$_2^-$ dumbbell turns clockwise in the $xy$ plane, whereas the other one turns anticlockwise (Table S2 in Ref. \cite{supplemental}). The averaged azimuth angle $\alpha$ for the two units of 40.2$^\circ$ is close to the value of 38.3$^\circ$ obtained at 100~K in the $I\bar{4}$ structure. Further, the tilting angles $\beta$ against the $c$ axis are similar for the two O$_2^-$ units but somewhat larger than the value at 100~K, $\sim$ 20$^\circ$ and $\sim$ 28$^\circ$ respectively. These structural reorientations alter substantially the shortest intermolecular O-O distances. As displayed in Fig. \ref{fig:super}(c) the shortest separations of the end oxygen atoms of superoxides form quadrilateral polygons in both polymorphs. In $I\bar{4}$ all edges are of the same lengths, the shorter ones among O3, the longer ones among O4. In $P\bar{4}$ this situation is changed, and moreover, the shorter bond lengths are splitting into two slightly different lengths, which again reflects symmetry breaking. As the active $\pi^*$ molecular orbitals have their lobes oriented perpendicular to the O-O bond axis, the structural transition at $T_{\rm s}$ should lead to a change in the orbital ordering pattern and affect the exchange interaction pathways.

The structural $I\bar{4}$ - $P\bar{4}$ transition at $T_{\rm s}$ is  dramatically reflected in the magnetic properties probed by  local probes of NMR and EPR. In the high-temperature tetragonal phase, the $^{87}$Rb NMR spectra have a characteristic quadrupolar powder lineshape with singularities of the central -1/2~$\leftrightarrow$~1/2 transition, split by $\sim 300$~kHz. As the temperature is lowered, the spectra further broaden and the singularities become less pronounced. At around  $T_{\rm s}$, there is a change in the $^{87}$Rb NMR lineshape. The quadrupolar singularities are no longer visible and the spectrum below $T_{\rm s}$ begins to adopt the shape of a very broad, almost featureless resonance.

The low-temperature broadening of the $^{87}$Rb NMR resonance suggests a change in the hyperfine coupling fields, whereas the accompanying disappearance of the quadrupole singularities is consistent with a structural distortion that affects the $^{87}$Rb quadrupole interaction via the electric field gradient. In order to throw some additional light on the transition at $T_{\rm s}$, we next turn to $^{87}$Rb spin-lattice and spin-spin relaxation measurements. In the high-temperature tetragonal phase, $1/T_1$  at first increases with decreasing temperature [Fig. \ref{fig:druga}(b)]. The spin-lattice relaxation rate is suddenly suppressed at $T_{\rm s}$  by a factor of $\sim 2$ within a 20~K interval and then it monotonically decreases with decreasing temperatures down to 5~K. $1/T_2$  has, within a factor of 5 in magnitude, an identical temperature dependence as $1/T_1$. This implies that both relaxation rates are governed by the same local-field fluctuations of purely magnetic origin with very short correlation times that dramatically change at $T_{\rm s}$.

Remarkably, the temperature dependence of the $^{87}$Rb $1/T_1$  in Rb$_4$O$_6$ is qualitatively very similar to that of the $^{133}$Cs $1/T_1$ in CsO$_2$ [inset in Fig. \ref{fig:druga}(b) and Ref. \citep{klanjvsek2015phonon}]. In CsO$_2$, the increase of $1/T_1$ with decreasing temperature between 220 and 80~K was ascribed to the dynamic modulation of the antiferromagnetic exchange coupling between nearest-neighbouring O$_2^-$ spins due to their thermal librations \citep{klanjvsek2015phonon}.
O$_2^-$ librations at $\sim 80$~K  suddenly freeze-out and trigger the structural transition from the tetragonal to the orthorhombic phase \citep{riyadi2012antiferromagnetic}. In the orthorhombic phase the O$_2^-$ dumbbells tilt and, as a result, the double degeneracy of O$_2^-$ molecular orbitals is lifted, thereby allowing orbital ordering that promotes the formation of antiferromagnetic spin chains to establish.
The resemblance of $^{87}$Rb $1/T_1$ in Rb$_4$O$_6$ to that of CsO$_2$ for temperatures above and around $T_{\rm s}$ leads us to the suggestion that  similar librations of  O$_2$ units in the tetragonal phase also determine the $^{87}$Rb $1/T_1$ for $T>T_{\rm s}$, while the freeze-out of the O$_2$ librations  at the structural transition at  $T_{\rm s}$ is accompanied by a concomitant  change in the oxygen dumbbell $\pi^*$ orbital order.

Plotting $1/T_1$ in logarithmic scale versus $1/T$  shows a straight line for $T \leq 70$~K [Fig. \ref{fig:druga}(c)]. Such thermally activated behavior of $1/T_1$ (and also of $1/T_2$) demonstrates the opening of a spin gap in the spectrum of low-energy spin excitations. By fitting the $1/T_1$ data below 70~K to $1/T_1 = A \exp (-\Delta _{\rm s}/k_{\rm B}T)$ we extract the size of the spin gap $\Delta _{\rm s}/k_{\rm B} =9$~K in the magnetic field of 9.39~T.
 We note that no evidence for the opening of a spin gap at low temperatures has been observed in $1/T_1$ data taken on the sister compound Cs$_4$O$_6$ \citep{arvcon2013influence}, which suggests a fundamental difference in their magnetic ground states.
It is also fundamentally different from CsO$_2$, where the orbital ordering leads to the formation of the antiferromagnetic chains of O$_2^-$ $S=1/2$ moments with the gapless Tomonaga-Luttinger-liquid state \citep{klanjvsek2015phonon, Kanflic2015}. The presence of a spin gap in the excitation spectrum of  Rb$_4$O$_6$ thus infers a new kind of low-temperature orbital order, which favorizes the antiferromagnetic coupling between pairs of O$_2^-$ spins.

In agreement with the NMR and structural data, we find a clear  anomaly at $T_{\rm s}$ in the X-band EPR linewidth and in  the \textit{g}-factor   [Fig. \ref{fig:tretja}(b) and (c)]. The latter is particularly dramatic, as the \textit{g}-factor increases from 1.8714 to the low-temperature value of 1.9997 in a narrow temperature interval of about 15~K. The components of the \textit{g}-factor tensor explicitly depend on the splitting between $\pi^*$ orbitals \cite{arvcon2013influence,kanzig1959paramagnetic}, so the change in the \textit{g}-factor at $T_{\rm s}$ is consistent with the structural $I\bar{4}$ - $P\bar{4}$ transition where the tilting of O$_2^-$ units changes. Moreover, if the low-temperature orbital order indeed affects the exchange coupling between the nearest neighboring O$_2^-$ groups, then also the EPR linewidth, which is in the exchange-narrowing limit given by $\Delta B \approx M_2/J$ \cite{bencini2012epr} (here $M_2$ is the second moment of the magnetic anisotropic interactions), is expected to change too. Therefore, the narrowing of the EPR signal from $\Delta B = 600$~mT at 95~K to $\Delta B=450$~mT at 80~K is yet another evidence of the correlation between structural and orbital order on one side and the spin state emerging from the modification of  exchange coupling constants at $T_{\rm s}$ on the other.

\subsection{Coupling of the magnetic properties to the orbital order below $T_{\rm s}$}

Below $T_{\rm s}$, the molar magnetic susceptibility $\chi_{\rm m} (T)$  of Rb$_4$O$_6$ monotonically increases with decreasing temperature [Fig. \ref{fig:tretja}(d)] and shows a pronounced maximum at $T_{\rm max} = 15$~K, below which it is rapidly suppressed. At the lowest temperatures, a Curie upturn dominates $\chi_{\rm m} (T)$, which can be attributed to unpaired O$_2^-$ ions due to the residual charge and orbitally disordered cubic phase of Rb$_4$O$_6$. This is corroborated by an increased Curie tail observed after the more rapid cooling down to 2~K, where a larger fraction of the cubic phase is frozen-in (Fig. S5 in Ref. \cite{supplemental}).

The integrated X-band EPR signal intensity, $\chi_{\rm EPR} (T)$ closely mimics $\chi_{\rm m} (T)$, i.e., it shows a pronounced maximum at $T_{\rm max}$, below which the EPR signal rapidly disappears due to the existence of a spin gap [Fig. \ref{fig:tretja}(d)]. At the lowest temperatures, only  a significantly narrower signal with a small axially-symmetric $g-$factor anisotropy ($g_{\perp} = 1.9757$ and $g_{\parallel}=2.311$) is observed [Fig. \ref{fig:tretja}(a) and Fig. S6 in Ref. \cite{supplemental} for the lineshape fit].
 In a control X-band EPR experiment using a quench cooling protocol from $400$~K, we measured the EPR signal of the quenched cubic phase alone (Fig. S7 in Ref. \cite{supplemental}). This signal, whose intensity nicely follows the Curie-like temperature dependence (Fig. S8 in Ref. \cite{supplemental}), is by a factor of $\sim 100$ broader than the  $g-$factor broadened low-temperature signal in the slow-cooling experiments and thus cannot explain the peculiar   X-band EPR residual  response at low temperatures. Alternatively, the origin of the  unpaired O$_2^-$ moments that contribute to such signal may be  due to some structural fluctuations with the fourth charge localizing on the "wrong" O$_2$ unit or even its delocalisation over several O$_2$ units around such defect. A possible realization of such structural fluctuation would be a defect that includes O$_2$ unit, which has not flipped from its original cubic orientation during the cubic-to-tetragonal transition. In this respect,  the presence of  structural fluctuations deep in the Verwey charge ordered phase may be reminiscent of "trimerons" in the canonical Verwey system Fe$_3$O$_4$ \cite{senn2012charge}.

We note that  neither the low-temperature maximum in  $\chi_{\rm m}(T)$ nor the anomalies at higher temperatures were apparent in earlier studies of the magnetic properties of Rb$_4$O$_6$ \cite{winterlik2009exotic}. As in the time of these studies the structural transitions were still unknown, possibly quite rapid cooling procedures were applied and thus rather the cubic or mixtures of the cubic and tetragonal phase were investigated at low temperatures. This rationalizes also the irreversibility in the $\chi _{\rm m}(T)$ curves which was previously attributed to frustration effects \cite{winterlik2009exotic} and underpins the crucial role played by the orbital order for the low-temperature spin state.

\begin{figure*}[t]
\includegraphics[width=\linewidth]{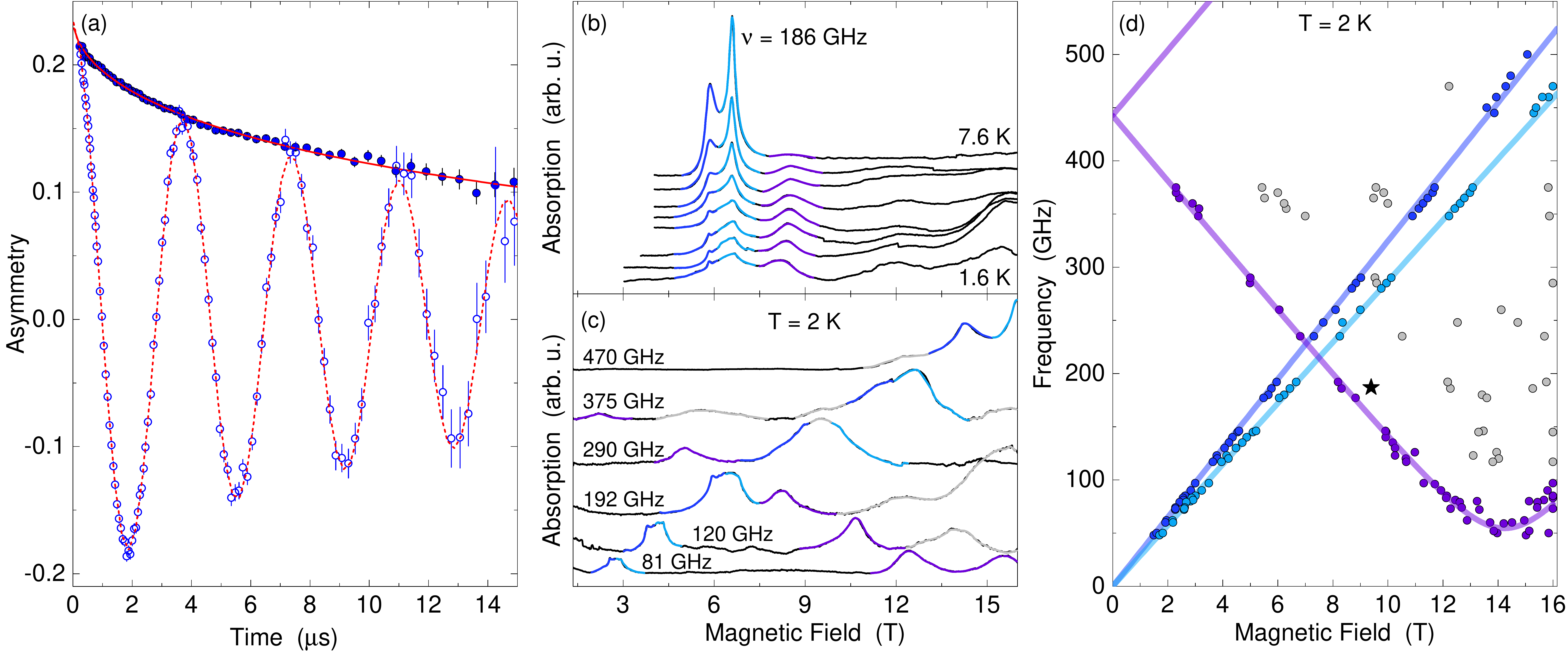}
\caption{(a) The zero-field (solid blue circles) and transverse-field (open blue circles) $\mu$SR asymmetry $A(t)$ of Rb$_4$O$_6$ powder measured at $T=1.6$~K. Note the complete absence of  oscillations in $A_{\rm ZF}(t)$ in zero applied field, which thus rules out any static order of O$_2^-$ moments down to 1.6~K. The solid red line is a fit to a stretched-exponential relaxation function $A_{\rm ZF}(t) = A_0 \exp \left[ - (\lambda t)^\beta\right]$ with the initial asymmetry $A_0=0.238$ determined in the complementary experiment in the transferred field geometry, relaxation rate $\lambda = 0.046~\mu$s$^{-1}$ and the stretching exponent $\beta = 0.54$. (b) Low temperature high-field EPR spectra measured at 186 GHz (from top to bottom at $T= 7.6$, 5.5, 4.6, 3.7, 2.8, 2.3, 1.9 and 1.6~K). The peaks are color-coded to match the modes in the resonance frequency-field diagram -- dark and light blue are the conventional EPR lines and purple lines due to the singlet-triplet (ST) transition mode. (c) High-field EPR spectra at 2~K measured at different frequencies. The peaks are color-coded in the same fashion as in (b), with the addition of light gray matching other observed transitions. (d) The resonance frequency-field diagram of Rb$_4$O$_6$ measured at 2~K. The dark and light blue circles show the conventional linear dependence characteristic of  EPR transitions within the triplet states, the purple circles show the singlet-to-triplet transition modes and the light gray circles are other observed modes of triplet or possibly quintet bound states origin. The black star is the spin gap value determined in NMR measurements.}
\label{fig:cetrta}
\end{figure*}

In contrast to the pure superoxides  CsO$_2$ and RbO$_2$ in which O$_2^-$ spins ultimately undergo three-dimensional antiferromagnetic order near 10~K \cite{klanjvsek2015phonon, Kanflic2015} and 15~K \cite{astuti19}, respectively, there are no indications for long-range magnetic order from the neutron diffraction patterns of the  Cs$_4$O$_6$  \cite{adler2018verwey, colman2019} and Rb$_4$O$_6$ [Fig. \ref{fig:super}(a)]. The absence of long-range magnetic order in Rb$_4$O$_6$ is here unambiguously confirmed by muon spin relaxation ($\mu$SR) experiments down to 1.6~K. The $\mu$SR asymmetry $A(t)$ measured at $T=1.6$~K in an applied transverse field (TF) of $ B_{\rm TF} = 2.0$~mT oscillates with the expected full initial asymmetry $A_0 = 0.238$ and frequency $\nu = 273.6$~kHz. This frequency corresponds well to the applied field ($\nu_{\rm TF} = \gamma_\mu  B_{\rm TF}/2\pi$, here $\gamma_\mu = 2\pi \cdot 135.5$~MHz/T is the muon gyromagnetic ratio), ruling out the presence of any ordered static internal fields. The absence of static internal fields is also revealed in zero-field (ZF) measurements, where the $\mu$SR asymmetry
exhibits a stretched-exponential type of decay [Fig. \ref{fig:cetrta}(a)] due to the fluctuating O$_2^-$ spins in the vicinity of the muon stopping site. Most importantly, we find  no oscillations in $A(t)$ that would  suggest long-range magnetic order. Moreover, the absence of a Kubo-Toyabe-like dip \cite{schenck1985muon} in $A(t)$  up to 15~$\mu$s limits any disordered static local fields at the muon stopping sites to values below $\sim$0.1~mT and thus corroborates a dynamic nature of the O$_2^-$ spins even at the lowest temperatures. The lack of any static magnetic order in the alkali sesquioxides is an important difference to the canonical Verwey system Fe$_3$O$_4$.

The low-temperature  $\chi_{\rm m} (T)$ of Rb$_4$O$_6$ is in striking contrast to that of Cs$_4$O$_6$, for which a Curie-Weiss-like $\chi_{\rm m} (T)$ was reported down to 2 K \cite{adler2018verwey}. This must originate from the difference in the orbital order as the second structural transition at $T_{\rm s}$ is absent in  Cs$_4$O$_6$.  The broad maximum in $\chi_{\rm m} (T)$ and an activated type behavior of $1/T_1$ are signatures of a gapped low-dimensional magnetism in Rb$_4$O$_6$. The simplest model that explains these findings is that of isolated dimers forming a singlet ground state. In this case, the molar magnetic susceptibility is given by the Bleaney-Bowers (BB) expression \citep{bleaney1952anomalous}
\begin{equation}
\chi_{\rm m} (T) = \dfrac{8C_{1/2}}{ T \left( 3 + \exp (J / k_B T) \right) }\, .
\label{BBeq}
\end{equation}
Here, $J$ is the intra-dimer exchange coupling constant and $C_{1/2}=N_{\rm A}\mu_0 \mu_{\rm B}^2g^2/(4k_{\rm B})$ is the Curie constant for $S=1/2$, where $N_{\rm A}$ is the Avogadro constant, $\mu_0 $ is the vacuum permeability, $ \mu_{\rm B}$ is the Bohr magneton and $ k_{\rm B}$ is the Boltzmann constant. The main features of the $\chi_{\rm m} (T)$ curve are reasonably well reproduced by this model if a small Curie contribution  is also added [Fig. \ref{fig:tretja}(d)]. An unconstrained fit with Eq. (\ref{BBeq}) yields $J/k_{\rm B}=25 \pm 0.3$~K and the effective magnetic moment $\mu_{\rm eff} = \mu_{\rm B}g\sqrt{3}/2 = 1.98~\mu_{\rm B}$. The latter is in good agreement with $\mu_{\rm eff} = 1.95~\mu_{\rm B}$ derived from the Curie-Weiss analysis of  $\chi_{\rm m} (T)$ between 50 and 120 K (Fig. S9 in Ref. \cite{supplemental}).
The complete disappearance of the main X-band EPR signal at the lowest temperatures fully complies with  a spin gap in the excitation spectrum.
Moreover, the maximum in  $\chi_{\rm EPR} (T)$ data is also reasonably well matched by Eq. (\ref{BBeq}), yielding a similar exchange coupling constant $J/k_{\rm B}=23 \pm 0.3$~K.
 All magnetic data thus point to the conclusion that   dimers of antiferromagnetically coupled O$_2^-$ spins are the main motif in the magnetic lattice of the  Rb$_4$O$_6$  low-temperature orbitally ordered $P\bar{4}$ structure.

\subsection{The low-temperature magnetic ground state}

To expose the microscopic nature of the magnetic ground state in the low-temperature orbitally-ordered state, i.e. for $T \leq T_{\rm s}$, we finally resort to multi-frequency EPR measurements.
The high resolution of the high-field EPR measurements enables us to resolve the splitting of the EPR spectra below 145~K  (Fig. S10 in \cite{supplemental}).  On further cooling, the splitting into two well distinct signals at  $g_1 = 2.32$ and $g_2 = 2.05$ becomes even more pronounced [Fig. \ref{fig:cetrta}(b)].  This indicates that the slowing down of the O$_2$ librations opens a path for the structural distortion that is responsible for the two inequivalent O$_2^-$ molecules in the unit cell, which is in excellent agreement with the low-temperature $P\bar{4}$ crystal structure obtained from the PND data [Fig. \ref{fig:structure}(c) and (d)]. The splitting of the EPR spectra already at temperatures above $T_{\rm s}$  may be a signature of structural fluctuations, similar to those found in Fe$_3$O$_4$ \cite{perversi2019co}.

Similar as in the X-band spectra, the high-field EPR signal intensity reaches a maximum at $T_{\rm max}$. However, as the intensity of the high-temperature EPR signals starts to decrease,  another signal (hereafter labeled as ST) positioned at the resonance field of 8.5~T and well separated from the two main peaks at $g \sim 2$ (i.e., at around 6~T) starts to gradually emerge below 8~K in the experiment conducted at a resonance frequency of 186~GHz, [Fig. \ref{fig:cetrta}(b)]. In sharp contrast to the paramagnetic $g \sim 2$ signals, the intensity of the ST signal increases with decreasing temperature.

The appearance of the ST signal in the spin-gapped phase at temperatures well below $T_{\rm max}$ is at first glance surprising and calls for additional frequency-field dependent EPR measurements. Such experiments were conducted at 2~K and are summarized in Figs. \ref{fig:cetrta}(c) and (d). The ST signal displays the opposite frequency-field dependence as compared to the conventional paramagnetic EPR signals at $g_1 = 2.32$ and $g_2 = 2.05$. Namely, with increasing frequency, the ST resonance field decreases. 
On the resonance frequency-field diagram [Fig. \ref{fig:cetrta}(d)], the ST signal shows almost perfect linear dependence that extrapolates to the resonance frequency of 443~GHz at zero field, which corresponds to a zero-field spin gap $\Delta _{\rm s}(0)/k_{\rm B} = 21$~K.  The spin gap of $\Delta _{\rm s}(9.39)/k_{\rm B} = 9$~K, determined in NMR measurements conducted  at 9.39~T [Fig. \ref{fig:druga} (c)], thus falls nicely on the EPR resonance frequency-field diagram. For fields larger that $\sim 13$~T, deviations from a simple linear dependence are observed and the resonance field even starts to increase with increasing frequency for fields larger than 14.25~T.

The high-field EPR measurements unambiguously confirm the existence of a zero-field spin gap $\Delta _{\rm s}(0)$ in the spectrum of low-energy spin excitations. In light of dominant antiferromagnetic interactions, the excitations could, in principle, either be magnons in an antiferromagnetically ordered state, or excitations from the singlet ground state to the triplet excited state. In the former case, the ST signal would be one of the antiferromagnetic resonance (AFMR) modes. However, the $\mu$SR data [Fig. \ref{fig:cetrta}(a)] unambiguously  rule out any antiferromagnetically ordered state and thus no AFMR signal is expected.
On the other hand, the ST signal shows all characteristics of the direct transition from the singlet to the triplet state: the temperature independent zero-field spin gap is given by the dominant antiferromagnetic exchange interaction between two O$_2^-$ spins while the increase of the ST signal intensity with decreasing temperature reflects the increase in the population of the singlet ground state with decreasing temperature. In this picture, the ST resonance frequency-field dependence for fields larger than 13~T, where the triplet branch approaches the singlet energy level, is due to a level anticrossing, which is a hallmark of the Dzyaloshinskii-Moriya (DM) interaction \citep{nojiri2003esr}. We stress that this interaction also provides the necessary mixing of quantum spin-states, which makes the direct singlet to triplet transition observable in EPR experiments.

In order to qualitatively describe the magnetic ground state of Rb$_4$O$_6$, we next attempt to fit the  EPR resonant field-resonant frequency data. In addition to the dominant antiferromagnetic exchange between pairs of O$_2^-$ spins, we take into account also the DM interaction as the leading magnetic anisotropy term. The latter is symmetry allowed since the center of inversion is absent at the center of each spin dimer in the low-temperature  $P\bar{4}$  structure. The complete Hamiltonian for a spin dimer then reads
\begin{equation} H= J \mathbf{S}_1 \cdot \mathbf{S}_2 + \mathbf{D} \cdot \mathbf{S}_1 \times \mathbf{S}_2 + g \mu _{\rm B} (\mathbf{S}_1 + \mathbf{S}_2 )\cdot \textbf{B} \, . \label{Hamilt}
\end{equation} Here, $\mathbf{D}$ is the vector of the DM interaction, while the third term accounts for the Zeeman interaction. This model is used to calculate the resonance frequency-field diagram and the best agreement with the experimental data gives $J/k_{\rm B}=21$~K and $D=2.5$~K [Fig. \ref{fig:cetrta}(d)]. The ratio $D/J= 0.12$ is surprisingly large for a light element $\pi$-electron system and is for instance of a similar magnitude as typically found in Cu$^{2+}$ systems \cite{nojiri2003esr,zorko2008dzyaloshinsky,zorko2013dzyaloshinsky}.

Although this simple model manages to explain the main features of $\chi_{\rm m}(T)$, $^{87}$Rb NMR and EPR data there are still some details, which may require further weaker terms in Eq. \ref{Hamilt}. Specifically, the  dependence of $\chi_{\rm m} (T)$ around $T_{\rm max}$ in the magnetic as well as in the EPR susceptibilities shows small deviation from the model predictions.
Even when the DM interaction is included in the model for $\chi_{\rm m}(T)$ \citep{stoll2006easyspin}, it shows only marginal deviation for temperatures below 2~K [Fig. \ref{fig:tretja}(d)]. More importantly, the high-field EPR data show many more peaks, especially at high fields [Fig. \ref{fig:cetrta}(b), (c) and (d)], which are also not predicted for the model of isolated dimers. Finally, it is difficult to comprehend that isolated spin dimers would form by the observed small structural change of the otherwise high-temperature tetragonal phase with three-dimensional (3D) pyrochlore  exchange network of O$_2^-$ sites \cite{colman2019}.  For these reasons we suggest that dimers of O$_2^-$ spins are in fact weakly coupled. If this is the case, then the additional EPR lines observed in the high-field experiments may represent triplet bound states, similar as found in SrCu$_2$(BO$_3$)$_2$ \citep{nojiri2003esr}, which is an archetypal two-dimensional (2D) lattice of interacting dimers. More detailed high-field data in the low-temperature phase is needed, in order to fully understand how the O$_2^-$ dimers interact and form the intricate singlet ground state in Rb$_4$O$_6$.

\section{Discussion and conclusions}

Similarly as 3$d$-based transition metal compounds, molecule-based alkali sesquioxides $A_4$O$_6$ and superoxides $A$O$_2$ with partially filled $\pi^*$ molecular orbitals are a versatile class of compounds for studying strongly correlated electron physics in the presence of orbital degeneracy. In this work we have demonstrated the strong entanglement between spin, charge, orbital, and lattice degrees of freedom in the anionic mixed-valence compound Rb$_4$O$_6$. Just like its sister compound Cs$_4$O$_6$ studied before \citep{adler2018verwey}, Rb$_4$O$_6$ features a Verwey-type charge-ordering transition near 290~K as confirmed here from structural and transport studies. EPR and NMR spectra  evidence charge localization on their respective time-scales of $10^{-10}$~s and $10^{-6}$~s. The first-order transition from the charge disordered cubic to the charge-ordered tetragonal structures involves a reorientation of the anionic O$_2$ molecular dumbbells and leads to a tilting of the paramagnetic O$_2^-$ units versus the $c$-axis. On the other hand, the low-temperature magnetic and structural properties  deep in the charge-ordered state of the two Rb- and Cs-based sister compounds are fundamentally different. While the magnetic susceptibility of Cs$_4$O$_6$ monotonically increases with decreasing temperature in the absence of any further structural changes \cite{adler2018verwey,arvcon2013influence}, Rb$_4$O$_6$ shows a pronounced maximum in $\chi_{\rm m}(T)$ as well as in $\chi_{\rm EPR}(T)$ at $T_{\rm max}$ $\approx$ 15 K as a result of dominant antiferromagnetic interactions within pairs of O$_2^-$ spins. The roots of this difference can be traced back to the differences in the orbital ordering patterns. The low-temperature crystal structure derived from the powder neutron diffraction data, the $^{87}$Rb NMR spectra and spin-lattice relaxation rates, as well as a $g$-factor anomaly in the EPR data provide consistent evidence for emergence of a peculiar type of $\pi^*$ orbital ordering at $T_{\rm s}$ = 92~K, which enables the dominant antiferromagnetic exchange within pairs of O$_2^-$ moments. We stress that the exchange interactions between nearest-neighboring O$_2^-$ spins are known to strongly depend on the precise orbital ordering \cite{riyadi2012antiferromagnetic} and thus provide a necessary link between the spin and orbital degrees of freedom.

The importance of the orbital ordering in the charge-ordered state in Rb$_4$O$_6$ is evident from our density-functional-theory (DFT) calculations. These calculations for the  high-temperature tetragonal structure (space group $I\bar{4}$, Ref. \cite{sans2014structural}) give the Wannier orbitals and the related hopping integrals between oxygen atoms on the nearest-neighboring dumbbell units. The resulting exchange-coupling network appears to be highly three-dimensional (Table S4 in Ref. \cite{supplemental}), without clear preference for the formation of the antiferromagnetically coupled O$_2^-$ pairs. The formation of such an exchange network seems to be the case in Cs$_4$O$_6$ for all temperatures.

 On the other hand, Rb$_4$O$_6$ undergoes a structural transition at $T_{\rm s}$ to the  $P\bar{4}$ crystal structure, which has pronounced consequences for the underlying orbital order. The calculated partial density of states (PDOS) for selected O$_2^-$ molecules, labeled with numbers from 5 to 12 in Figs. \ref{fig:structure}(c) and (d), are shown in Fig. S11. They disclose a systematic pattern, where three of spin-up and spin-down states of the two $\pi^*$ orbitals are almost fully occupied leaving only one of these empty above the Fermi level. PDOS thus exhibit a clear preference for the alternating O$_2^-$ $\pi_x^*$ and $\pi_y^*$ orbital order  of O$_2^-$ units stacked along the tetragonal $c$ axis   [Fig. \ref{fig:structure}(c) and (d)].  What is remarkable for this structure is that the tilting together with the orbital ordering enables a direct hopping between oxygen atoms of the nearest neighbouring O$_2^-$ units with the strongest hopping between the oxygen atoms on  O$_2^{-}$  pairs $5-8$, $6-12$, $7-11$ and  $9-10$. We note that such hopping pattern is in general agreement with the structural data as  these pairs have  short end-to-end O-O distances [Fig. \ref{fig:super}(c)]. All these hopping integrals are within the precision of our computations nearly the same, i.e., $t^2=0.039$~eV$^2$ and therefore explain the formation of spin dimers as depicted in  Figs. \ref{fig:structure}(e)  and (f). Other direct hopping integrals  or  pathways that include hopping via O$_2^{2-}$ or Rb $p-$orbitals are smaller and thus may lead only to the weak exchange coupling between the spin dimers. In order to directly estimate the antiferromagnetic spin dimer exchange coupling, $J$, we next calculate the total energies of the two magnetic configurations: the one with the antiferromagnetic alignment of O$_2^{-}$ spins on the spin dimers $5-8$, $6-12$, $7-11$ and  $9-10$, and the one with the ferromagnetic alignment of all spins. By  comparing the relative energies of these two magnetic configurations (Table S5 in Ref. \cite{supplemental}), we compute the intra-dimer exchange $J=30.7$~K, which is in reasonable agreement with the experimental data [Figs. \ref{fig:tretja}(d) and \ref{fig:cetrta}(d)]. The small discrepancy between the experimental and DFT values  of $J$ may partly arise from the neglected inter-dimer exchange.

The observed coupling between the orbital and spin degrees of freedom may be in Rb$_4$O$_6$ facilitated through two different physical mechanisms. The first mechanism involves  a Kugel-Khomskii-type behavior \cite{kugel73}, where the orbital ordering is driven by exchange interactions. We note that the Kugel-Khomskii model was successfully applied to cuprates with degenerate $e_g$ orbitals, such as for example KCuF$_3$ or K$_2$CuF$_4$ \cite{oles2017orbital}, but rarely to $\pi$ electron systems. Alternatively, the compressed lattice in Rb$_4$O$_6$ compared to Cs$_4$O$_6$ could trigger an instability of the high-temperature tetragonal $I\bar{4}$ structure via an additional rotation of the O$_2^-$ units and result in a lowering of the structural symmetry to $P\bar{4}$  and  concomitant increasing in the crystal-field splitting between the $\pi^*$ orbitals. The magnetic interactions giving rise to spin dimerization then would be rather a consequence of the structural instability. We stress that the relative importance of the Kugel-Khomskii-type exchange and Jahn-Teller-like electron-lattice coupling is currently an important subject for understanding orbital-ordering phenomena \cite{pavarini08,pavarini10}.

The interplay between the Kugel-Khomskii-type superexchange, crystal-field splitting, spin-orbit coupling (SOC), and the resulting orbital-ordering patterns was theoretically analyzed for the monovalent superoxide KO$_2$ \cite{kim14}, where a series of structural tranitions involving different types and degrees of tilting of the O$_2^-$ units was observed. For the small tilting angles of the O$_2^-$ units versus the $c$ axis in the basic tetragonal structure the Kugel-Khomskii mechanism together with SOC dominates, whereas for larger tilting angles the crystal-field splitting becomes more important. For intermediate angles competitition between the various interactions occurs. Our structural studies reveal a quite large tilting angle of $\sim 20^\circ$ for Rb$_4$O$_6$ in the $I\bar{4}$ and $\sim 28^\circ$ in the $P\bar{4}$ structure, which is even enhanced compared to that of $\sim 17^\circ$ in Cs$_4$O$_6$. Thus, based on Ref. \cite{kim14}, the crystal-field splitting of the degenerate $\pi^*$ molecular orbitals is expected to stabilize the tetragonal $I\bar{4}$ charge-ordered structures in both, Rb$_4$O$_6$  and Cs$_4$O$_6$. The larger tilting angle indicates larger orbital splitting in Rb$_4$O$_6$ than in Cs$_4$O$_6$ which is in agreement with the persistence of the tetragonal structure to higher temperatures in Rb$_4$O$_6$. The final remaining question is about the driving force of the orbital ordering transition at $T_{\rm s}= 92$~K. Here, as the temperature is significantly lower, the Kugel-Khomskii mechanism could play an important role and open the path for an additional structural adaptation by rotation of the O$_2^-$ units and spin dimer formation. A theoretical modeling of the sesquioxides incorporating also the hopping terms due to the mixed-valence situation is highly desirable to resolve the nature of the different structural transitions and especially to clarify the detailed role of Jahn-Teller-like structural distortions and Kugel-Khomskii-type exchange in orbital ordering of mixed-valence strongly correlated electron systems with orbital degeneracy  \cite{pavarini08,pavarini10}.

The present comprehensive study of the anionic $\pi$-electron mixed-valence compound Rb$_4$O$_6$  unambiguously demonstrates, how lattice, charge, orbital and spin degrees of freedom couple and tune the ground state in alkali sesquioxides.  In this respect alkali sesquioxides  are reminiscent of many notable mixed-valence $d$-electron systems, such as cuprates, manganites and Fe$_3$O$_4$ or more complex $p$-electron systems like fullerides \cite{klupp12}. In Fe$_3$O$_4$, the charge and orbital ordering leads to an interesting magnetic ground state of coupled trimerons \citep{senn2012charge,perversi2019co} -- a highly complex spin formation deep in the Verwey phase. The low-temperature EPR spectra in Rb$_4$O$_6$ may yield evidences for  analogous structural/orbital/spin fluctuations inherent to charge-ordered $\pi$ electron alkali sesquioxides. In these light element molecular solids, the natural frequencies of spin excitations are lowered and thus become accessible to most spectroscopic methods. Studying mixed-valence alkali sesquioxides may thus in the future provide deeper understanding of the complex excitations in Verwey-type compounds. Finally, the lack of long-range magnetic order as apparent from the present ZF $\mu$SR measurements down to 1.6 K suggests that also the quantum magnetism of this class of compounds deserves further attention.

\section*{Acknowledgements}
 We thank Patrick Merz for preparation of the samples, Ralf Koban and Walter Schnelle for the magnetization measurements, and Jörg Daniels for assistance with structural drawings. Ininitial stages of the project were supported by the European Union FP7-NMP-2011-EU-Japan project LEMSUPER under contract no. 283214. This work was supported by the Deutsche Forschungsgemeinschaft (DFG), through ZV 6/2-2, and  by the HLD at HZDR, member of the European Magnetic Field Laboratory (EMFL). D.A. acknowledges financial support from the Slovenian Research Agency (Core Research Funding No. P1-0125
and projects No. J1-9145 and N1-0052). Experiments at the ISIS Neutron and Muon
Source were supported by a beamtime allocation RB1820363 from the Science and Technology Facility Council. These measurements are available in Ref. \cite{Arcon2019ISIS}.

\section*{Methods}

Rb$_4$O$_6$ was prepared by thermal decomposition of the superoxide RbO$_2$ as described in Ref. \cite{merz17}. The purity of the samples was checked by laboratory powder x-ray diffraction (XRD) measurements using monochromatic CuK$_{\alpha 1}$ radiation. Due to the extreme air and moisture sensitivity of alkali metal oxides, all sample handlings were carried out under carefully controlled inert atmospheres.

Powder neutron diffraction (PND) experiments on  Rb$_4$O$_6$ were carried out on the instruments E6 and E9 at the BER II reactor of the Helmholtz-Zentrum Berlin. These instruments use a pyrolytic graphite (PG) and a Ge-monochromator selecting the neutron wavelengths  $\lambda = 2.426$~\AA\ (E6) and $\lambda = 1.7985$~\AA\ (E9), respectively. For the PND experiments, about 4 g of Rb$_4$O$_6$ powder, which was sealed in a thin-walled quartz tube and then placed in a vanadium can (6 mm diameter), was used. Powder data of Rb$_4$O$_6$ were collected on E6 between the diffraction angles 8 and 136.5$^\circ$  and on E9 between 8 and 141.8$^\circ$. On the instrument E9 full data sets of Rb$_4$O$_6$ were collected at 3, 100 and 400 K to determine precisely the positional parameters of the cubic and tetragonal structures, respectively. 
The refinements of the crystal structures were carried out with the program FullProf \cite{FullProf} with the nuclear scattering lengths $b$(O) = 5.805 fm and $b$(Rb) = 7.09 fm \cite{sears95}. 

On the instrument E6 we have investigated in detail the structural changes as a function of temperature in the range from 3 to 427 K. For all experiments we have used an Orange Cryofurnace (AS Scientific Products Ltd., Abingdon, GB). Considering that the structural and electronic properties of alkali sesquioxides sensitively depend on the thermal protocol \cite{arvcon2013influence} two different types of PND experiments were performed. In the first one the virgin sample was heated to above 400 K and then subsequently cooled down slowly. Between 250 and 380 K the cooling rates were kept to 1 K/min. Finally, the sample was cooled down to base temperature (2 K) and the sample was subsequently heated up again to 400 K. At selected temperatures PND patterns were collected with a measurement time of about 35 min at E6 and of 20 hrs at E9. In a second fast cooling experiment the sample was quenched into liquid nitrogen from 400 down to 80 K with an average cooling rate of about 40 K per minute. Then it was further cooled down to 3 K in the cryostat with a cooling rate of 2-3 K per minute. Subsequently the sample was heated up again, where the set temperatures for data collection were reached within 10 minutes.

Impedance measurements were performed on Rb$_4$O$_6$ pellets having a diameter of 6 mm and a thickness of around 1 mm following the experimental and data evaluation procedures described in Ref. \cite{adler2018verwey}. The measurements were carried out in a temperature range of 363 to 200 K in steps of 5 K during slow cooling down to 200 K and the subsequent also on heating using the same temperature step.

Magnetization measurements on powders of Rb$_4$O$_6$ sealed in a Suprasil quartz tube were performed with an MPMS3 (Quantum Design) magnetometer in field-cooling and field-heating modes in a magnetic field of 0.1 T. Starting from 400 K where the sample is entirely in the cubic phase, the sample was cooled with a rate of 2 K/min down to 2 K and subsequently heated back to 400 K at the same rate.

Zero-field (ZF) muon spin relaxation ($\mu$SR) experiments  were performed in an Oxford Instruments Variox cryostat on the MUSR instrument (ISIS, Rutherford Appleton Laboratory, United Kingdom) after slow cooling of the Rb$_4$O$_6$ sample to low temperatures. The sample was protected against the exposure to air in a home-made sample container.

Continuous wave X-band EPR measurements were done using a home-built spectrometer equipped with a Varian E-101 microwave bridge operating at 9.37~GHz, a Varian TEM104 dual cavity resonator, an Oxford Instruments ESR900 cryostat and an Oxford Instruments ITC503 temperature controller with temperature stability better than $\pm 0.05$K at all temperatures. For the purpose of the measurements, the sample was sealed in a Suprasil quartz tube (Wilmad-LabGlass, $4$~mm medium wall tube) under dynamic vacuum.

High-Field EPR measurements were done at the Dresden High Magnetic Field Laboratory (HLD) in the Helmholtz Zentrum Dresden Rossendorf. The experiments were performed using a multifrequency  transmission-type  EPR spectrometer equipped with a 16~T superconducting magnet in a Faraday configuration, similar to that  described in Ref. \cite{Zvyagin_INSR}. In our experiments, a set of VDI microwave sources  were used, allowing us to probe magnetic excitations in this material in the quasi-continuously covered frequency range from approximately 50 to 500 GHz. The thermal protocol included slow cooling to base temperature $T=1.6$~K. The sample was the same as in X-band EPR experiments.

NMR measurements of the $^{87}$Rb ($I=3/2$) frequency-swept spectra were done in a  4.7~T superconducting magnet. We used a standard solid-echo pulse sequence $(\pi/2)- \tau - (\pi/2)- \tau$ and  appropriate phase cycling. The radio-frequency pulse length was optimized to  $t_{\pi/2}=3.2$~$\mu$s and the interpulse-delay was set to $\tau$ = 50 $\mu$s. Typical repetition time was 20~ms. The complete wide-line polycrystalline NMR spectrum was obtained by summing the real part of individual spectra measured step by step at resonance frequencies separated by $\Delta\nu=50$~kHz. The spin-lattice and spin-spin relaxation rates  were measured using a 9.39~T magnet with the inversion-recovery technique. All NMR measurements were performed in the temperature range between 5 and 380~K using a slow cooling protocol.

DFT calculations were done by applying the Quantum Espresso
code \cite{Giannozzi:2009}. The exchange-correlation effects were calculated by
means of the generalized-gradient approximation \cite{perdew1996generalized}
with explicitly added \cite{cococcioni2005linear} Hubbard repulsion term
$U=4.1\,{\rm eV}$ for the oxygen atoms. The electron-ion interactions were
described with the
gauge-including-projected-augmented-waves (GIPAW)
pseudopotentials \cite{pickard2001all,DalCorso:2014} making us possible to calculate the NMR-based quantities.  The plane waves and the
charge-density cut-off parameters were set to $639\,{\rm eV}$ and $2543\,{\rm eV}$, respectively, whereas a $4\times 4\times 4$ mesh of $\mathbf{k}$-points \cite{Monkhorst:1976}
was used for the Brillouin-zone integration. The self-consistent criterion was the total-energy difference between the two subsequent calculations being less than
$10^{-6}\,{\rm Ry}$.
In order to initially estimate the exchange coupling between particular oxygen
dumbbell units, we determined the hopping integrals between the corresponding
maximally-localized Wannier orbitals, which resulted from the wannierization
performed with the Wannier90 tool \cite{Wannier90}.

%\bibliography{Refs_3p0}
%

\begingroup
\pagestyle{empty}
\cleardoublepage
\endgroup
\includepdf[pages=1]{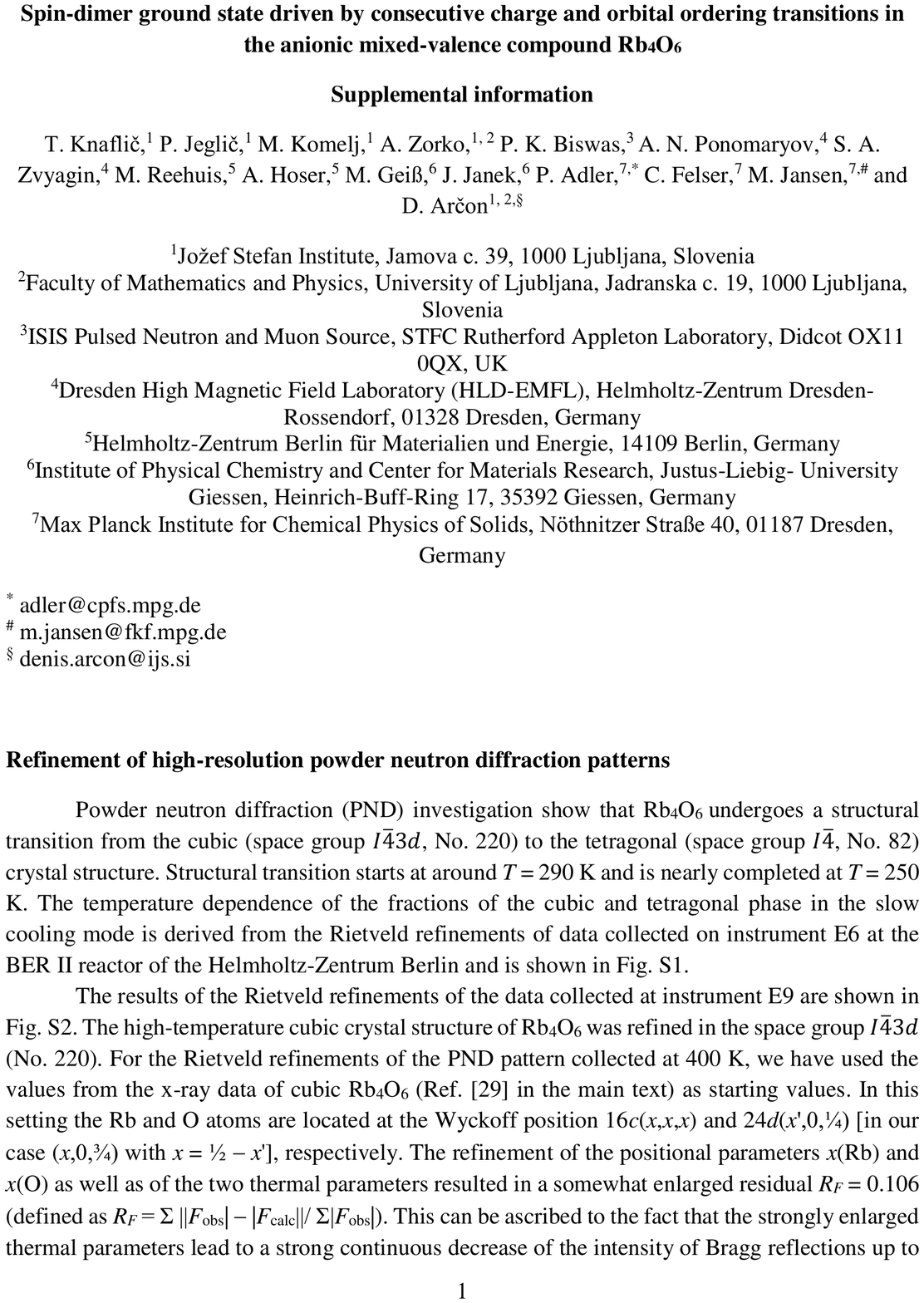}
\begingroup
\pagestyle{empty}
\cleardoublepage
\endgroup
\includepdf[pages=2]{Rb4O6-SM2.pdf}
\begingroup
\pagestyle{empty}
\cleardoublepage
\endgroup
\includepdf[pages=3]{Rb4O6-SM2.pdf}
\begingroup
\pagestyle{empty}
\cleardoublepage
\endgroup
\includepdf[pages=4]{Rb4O6-SM2.pdf}
\begingroup
\pagestyle{empty}
\cleardoublepage
\endgroup
\includepdf[pages=5]{Rb4O6-SM2.pdf}
\begingroup
\pagestyle{empty}
\cleardoublepage
\endgroup
\includepdf[pages=6]{Rb4O6-SM2.pdf}
\begingroup
\pagestyle{empty}
\cleardoublepage
\endgroup
\includepdf[pages=7]{Rb4O6-SM2.pdf}
\begingroup
\pagestyle{empty}
\cleardoublepage
\endgroup
\includepdf[pages=8]{Rb4O6-SM2.pdf}
\begingroup
\pagestyle{empty}
\cleardoublepage
\endgroup
\includepdf[pages=9]{Rb4O6-SM2.pdf}
\begingroup
\pagestyle{empty}
\cleardoublepage
\endgroup
\includepdf[pages=10]{Rb4O6-SM2.pdf}
\begingroup
\pagestyle{empty}
\cleardoublepage
\endgroup
\includepdf[pages=11]{Rb4O6-SM2.pdf}
\begingroup
\pagestyle{empty}
\cleardoublepage
\endgroup
\includepdf[pages=12]{Rb4O6-SM2.pdf}
\begingroup
\pagestyle{empty}
\cleardoublepage
\endgroup
\includepdf[pages=13]{Rb4O6-SM2.pdf}
\begingroup
\pagestyle{empty}
\cleardoublepage
\endgroup
\includepdf[pages=14]{Rb4O6-SM2.pdf}
\begingroup
\pagestyle{empty}
\cleardoublepage
\endgroup
\includepdf[pages=15]{Rb4O6-SM2.pdf}
\begingroup
\pagestyle{empty}
\cleardoublepage
\endgroup
\includepdf[pages=16]{Rb4O6-SM2.pdf}
\begingroup
\pagestyle{empty}
\cleardoublepage
\endgroup
\includepdf[pages=17]{Rb4O6-SM2.pdf}
\begingroup
\pagestyle{empty}
\cleardoublepage
\endgroup
\includepdf[pages=18]{Rb4O6-SM2.pdf}
\begingroup
\pagestyle{empty}
\cleardoublepage
\endgroup
\includepdf[pages=19]{Rb4O6-SM2.pdf}
\begingroup
\pagestyle{empty}
\cleardoublepage
\endgroup
\includepdf[pages=20]{Rb4O6-SM2.pdf}
\begingroup
\pagestyle{empty}
\cleardoublepage
\endgroup
\includepdf[pages=21]{Rb4O6-SM2.pdf}

\end{document}